\documentclass[twocolumn]{article}
\pdfoutput=1 
\usepackage[utf8]{inputenc}
\usepackage{authblk}
\title{Au$_{12}$@Au$_{30}$: Core-Shell Molecule Constituted of an Icosahedron and an Icosidodecahedron}
\author[a]{Bai Chunyuan}
\author[a]{Li Hongfei}
\author[a]{Xie Zun \thanks{Corresponding author: zxie@hebtu.edu.cn}}
\author[b]{Dong Yichen}
\author[c]{Liu Shulan}
\affil[a]{Department of Physics and Hebei Advanced Thin Film Laboratory, Hebei Normal University,}
\affil[b]{Department of Chemistry \& Environmental Science, Hebei University}
\affil[c]{College of Physical Science and Technology, Tangshan Normal University}

\date{June 2021}

\usepackage{natbib}
\usepackage{graphicx}
\usepackage{indentfirst} 
\usepackage{booktabs}
\usepackage{multirow} 
\usepackage{threeparttable}
\usepackage{endnotes}
\usepackage{multicol}
\usepackage{natbib}[number]

\begin{document}
\maketitle
\begin{abstract}
A stable core-shell structure with Ih symmetry, Au12@Au30, has been investigated by first-principles calculations. It is composed of an icosahedron core and an icosidodecahedron shell. The stability of the core-shell Au$_{42}$ structure is verified by vibrational frequency analysis and molecular dynamics NVT simulations. Both the frontier molecular orbitals and the spin density of states show obvious s-d hybridization characteristics. The adaptive natural density partitioning analysis demonstrate multi-center bonds, twenty 6-center $\sigma$ bonds ,and one 12-center $\sigma$ bond, which are of great importance for the core-shell structural stability. In this core-shell nanostructure, there are also a large number of one-center valence lone electron pairs with the characteristics of d-like orbitals, so that the proposed Au$_{12}$@Au$_{30}$ could be used in medicine and catalysis fields.
\end{abstract}
\setlength{\baselineskip}{1.5em} 
\setlength{\parskip}{2.0ex} 

\section{Introduction}
With the achievements of computing technology, more and more cluster structures composed of Au atoms have been discovered. These zero-dimensional structures can be widely applied in many fields, such as biology, medicine, nanotechnology, and chemistry catalysis of the reaction\citep{01},\citep{02},\citep{03},\citep{04},\citep{05},\citep{06}. Thus, the gold clusters have always been of interest. Most relativistic density functional theory (DFT) computations predicted the shape of Au$_n$ is a planar triangular valve, until n=7. When n $>$ 8, the system can form 3-dimensional structures\citep{07},\citep{08},\citep{09}. Li jun et al. discovered a regular tetrahedral cluster composed of 20 Au atoms in 2003\citep{10}, which has been researched until now\citep{11}. A fullerene structure of Au$_{32}$ was discovered by Mikael P. Johansson et al\citep{12} in 2004. It was discovered that the icosahedron Au$_{42}$ fullerene cage has excellent properties in the same year\citep{13}. Wang Jin lan et al. studied a series of  middle-size Au$_{n}$ clusters (n = 32, 38, 44, 50, 56) and found hollow cage configurations can compete energetically with their space-filling counterparts\citep{14}. In 2018, a kind of chiral symmetry fracture produced an I-Au$_{60}$ perfect gold shell with strange rigidity that entered people’s field of vision\citep{15}.

Besides the above hollow nanocage structure, some stable core-shell nanostructures have been discovered and explored one after another. From the past research, the ligand-protected metal clusters, Au$_{13}$@Au$_{42}$, had been found to have some applications in selective oxidation\citep{16}. An interesting thing is the most stable structure of its isomers is not this icosahedral pattern, but an amorphous form\citep{17},\citep{18}. In last year, a specific core-shell structure which is constituted of aluminum atoms has certain applications in blasting\citep{19}. In 2020, the core-shell Au@M (M=Pd, Pt) nanoparticle trimers synthesized was found to be able to achieve efficient conversion between light energy and chemical energy from S. Lee`s research\citep{20}. In the field of catalysis, a typical Au-Pd core-shell nanocrystals with near-surface alloy and single-layer Pd shell structure have excellent catalytic activity for hexavalent chromium conversion\citep{21} and some core-shell structures formed by Au and Pt have significantly enhanced catalytic activity for the electrooxidation of ethanol in alkaline media\citep{22}. Therefore, clusters with core-shell structures constituted of transition metals attracted our attention, and then we conceived and verified the existence of 42-atom core-shell structures. 

In this essay, we introduce the discovery process of this core-shell Au$_{12}$@Au$_{30}$ structure, prove its dynamic and thermal stability and also analyzed its internal bond formation and electronic structure. All calculations used are based on first principles, the details of the calculations are described in detail in the computation method section.

\section{Computation Method}
The calculations of the core-shell Au$_{42}$ were performed using the DMol$^{3}$ method\citep{23},\citep{24} based on spin-polarized density functional theory (DFT). The exchange-correlation interaction was treated within the generalized gradient approximation (GGA) using two different exchange-correlation functionals, Perdew-Burke-Ernzerhof (PBE)\citep{25} and Perdew-Wang 1991 (PW91)\citep{26}, to reduce uncertainties. In the electronic structure calculations, Double Numerical plus polarization basis sets was chosen here. Considering that the valence electrons of Au atoms (5d$^{10}$, 6s$^1$) have obvious relativistic effects, the DFT-based relativistic semi-nuclear pseudopotential (DSPP)\citep{27} was employed for the atomic core. In the structural optimization, without any symmetry constraints, the convergence criterion in the calculation process was set as the energy error within 1.0×10$^{-5}$ Ha, the force is 2.0×10$^{-3}$ Ha/Å and the atomic displacement is 5.0×10$^{-3}$ Å. 

The Born-Oppenheimer molecular dynamics (MD) simulations were carried out under the canonical ensemble (NVT). At init temperature , the total simulation duration was set to four cases 0.5ps, 1.0ps, 1.5ps, 2.0ps and the time step of each case is 1fs. The arithmetic average of the temperature values of all dynamic steps is recorded as the "equivalent"  temperature.

The bonding structure was analyzed by the adaptive natural density partitioning (AdNDP) method\citep{28} combined with the electronic localization function (ELF)\citep{29}. Here, we employed the PBE functional and the LANL2DZ pseudo core potential and basis-set\citep{30} as implemented in the Gaussian 09 package\citep{31}. The visualization of the calculated results was realized by Multiwfn software\citep{32} and Visual Molecular Dynamics (VMD) software\citep{33}.

\section{Result and Discussion}
The construction process of Au$_{12}$@Au$_{30}$ is shown in Figure 1. Au${_12}$@Au$_{30}$ is composed of an icosahedron (twenty triangles) core and an icosidodecahedron (twelve pentagons and twenty triangles) shell, which has the Ih symmetry, its side view and top view are shown in Figure 1(b, c). Obviously, there are only two kinds of unequal atoms in this system, one is the shell atom (denoted as Au$^1$), the other is the core atom (denoted as Au$^2$). Table S1 lists the cartesian coordinates of Au$_{12}$@Au$_{30}$.
\begin{figure}[h!]
\centering
\includegraphics[scale = 0.3]{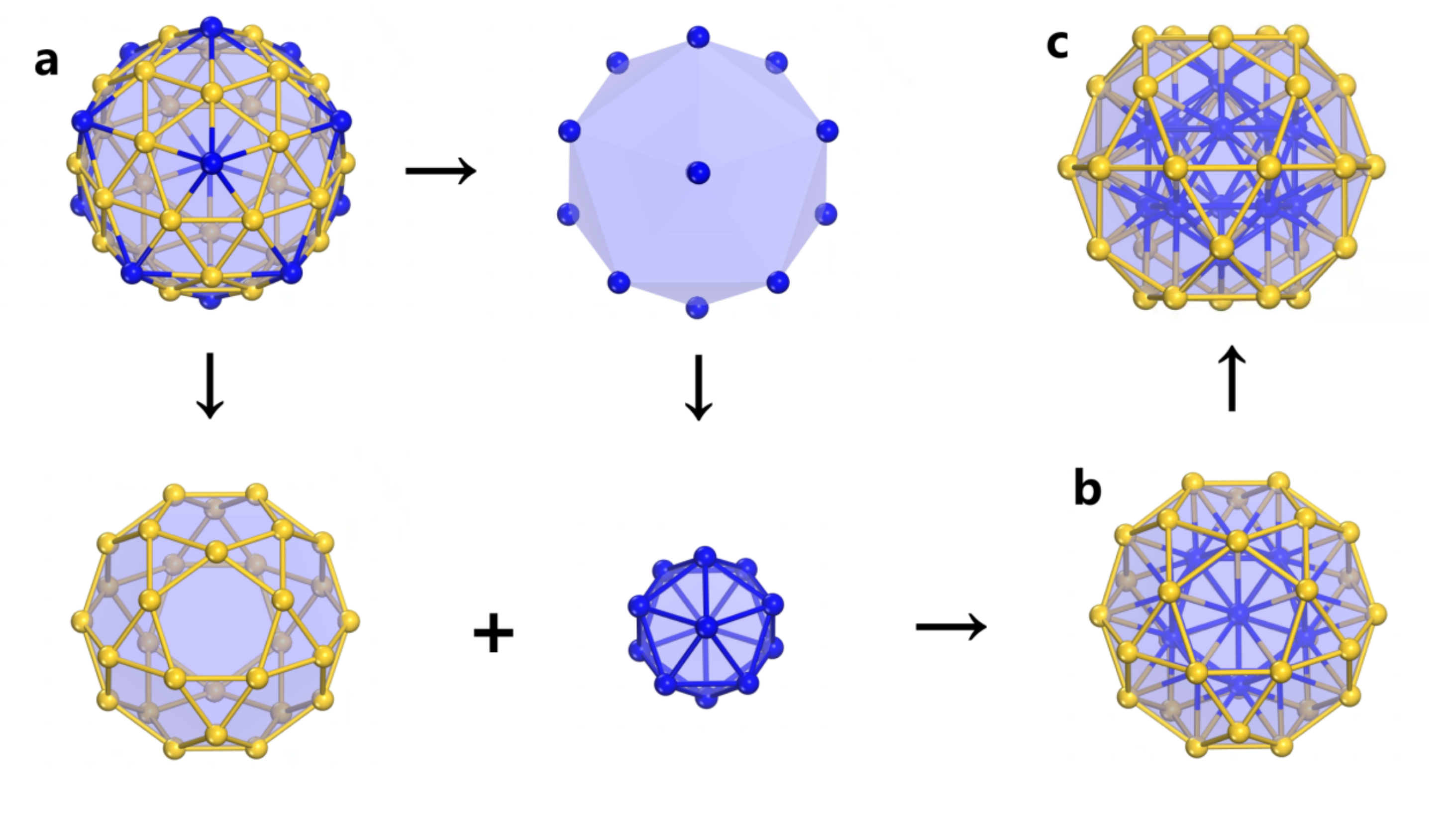}
\caption{The construction of the core-cell structure of Au$_{12}$@Au$_{30}$. The top view of the cage-like Au$_{42}$ (a) and the core-shell Au$_{42}$ (b). These 12-blue atoms in (a) collapse inward to form the blue core in (b). The side view (c) of Au$_{12}$@Au$_{30}$.}
\label{01}
\end{figure}

Some calculated properties of the stable core-shell structure obtained with the PW91/PBE exchange-correlation functional are summarized in Table 1. After energy minimization without any symmetry constrict, the Au$_{42}$ cluster retains the core-shell structure and the Ih symmetry, whether the PBE or the PW91 functional is used. At the temperature of 10K, the Raman spectrum was obtained and shown in Figure 2 and the wavelength of the irradiating light was used in the simulation is 514.50nm. From Figure 2, the vibration frequencies of all atoms in this structure are real, which proves its dynamic stability. For MD simulations in the NVT ensemble, the core-shell structure can still maintain the original shape at a temperature of about 450K. The simulations of 450K with different durations are shown in Figure S2. When the system is at a higher temperature (such as 500K, 5ps), the interaction between Au$^{2}$ atoms is destroyed firstly, causing the core atoms to move toward the direction of the cluster shell (shown in Figure S3). Why could not this system maintain its original structure at higher temperatures? We think that the reason is that the bonding in the system are all metallic bonds, the quantitative analysis can be reflected in the ELF analysis (shown in Figure 3) mentioned below. 
\begin{figure}[h!]
\centering
\includegraphics[scale = 0.3]{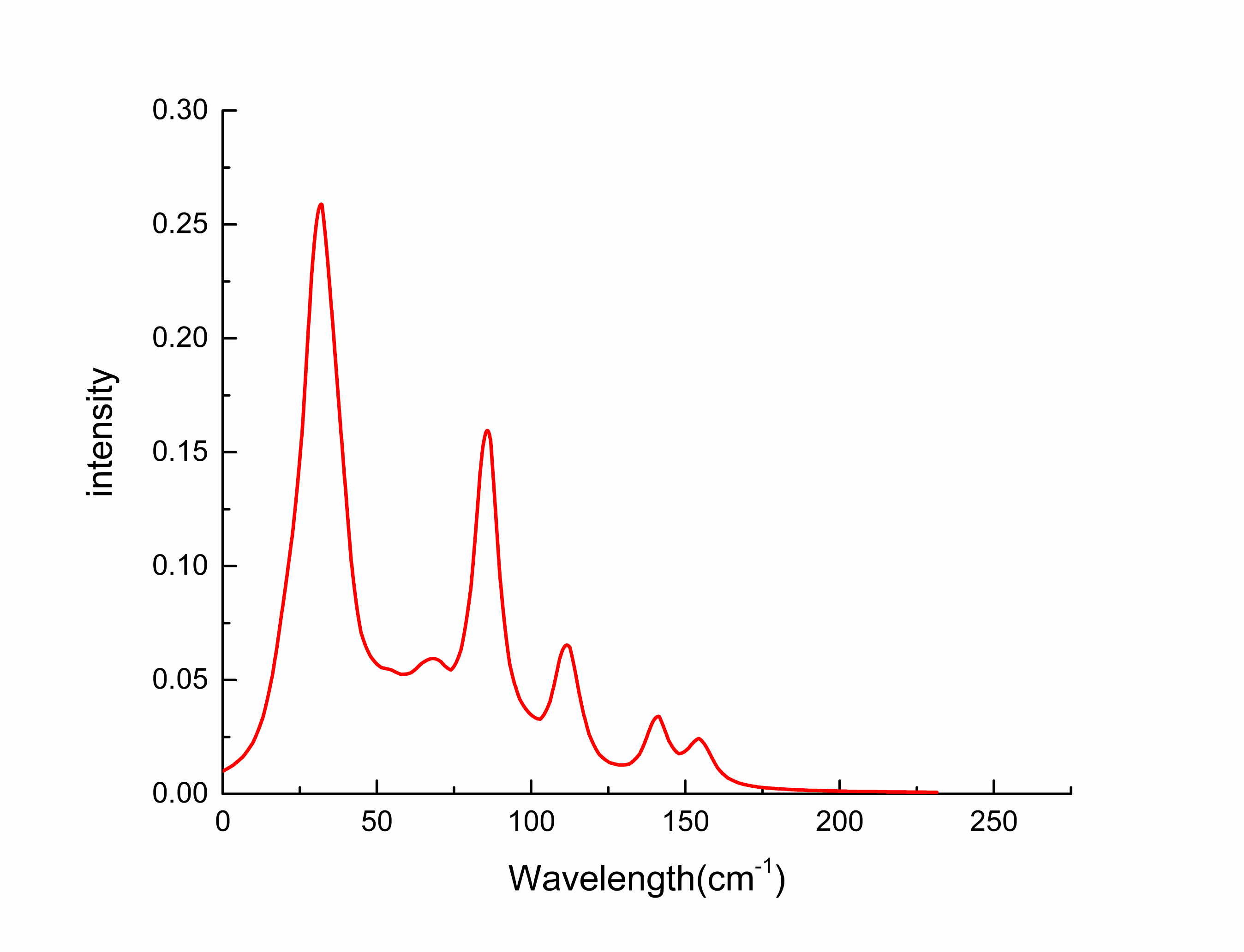}
\caption{The Raman Spectrum of the core-shell structure of Au$_{42}$ at 10K, the wavelength of light is 514.50nm and the Lorentzian smearing is 10.00 cm$^{-1}$(Calculated by PBE functional).}
\label{02}
\end{figure}

\begin{table*}[t]
\centering
\renewcommand\tabcolsep{3.5pt}
\begin{threeparttable}
\caption{Calculated the properties of Au$_{12}$@Au$_{30}$ at the GGA/PBE and GGA/PW91 levels.\tnote{1}}
\label{table1}
\begin{tabular}{|c|c|c|c|c|c|c|c|c|c|c|}
\hline
 & \multirow{2}{*}{Sym} & \multirow{2}{*}{Gap} & \multirow{2}{*}{d$_{c}$} &  \multirow{2}{*}{d$_{c-s}$}& \multirow{2}{*}{d$_{s}$} & \multirow{2}{*}{E$_{b}$} & \multicolumn{2}{|c|}{Freq} & \multicolumn{2}{|c|}{Hirshfeld}\\
\cline{8-11}
& & & & & & & f$_{max}$ & f$_{min}$ & Au$^{1}$ & Au$^{2}$ \\
\hline
PBE & Ih & 0.624 & 2.854 & 2.855 & 2.955 & 2.23 & 152.82 & 18.51 & -0.0143 & 0.0353\\
\hline
PW91 & Ih & 0.624 & 2.854 & 2.855 & 2.954 & 2.24 & 139.60 & 4.97 & -0.0144 & 0.0361\\
\hline
\end{tabular}
\label{table_MAP}
     \begin{tablenotes}
       \footnotesize
       \item[1] Point group symmetry (sym), HOMO-LUMO gap (Gap)(eV), the bond length (Å) between core atoms (d$_c$), between shell atoms (d$_s$), between core atoms and shell atoms (d$_{c-s}$), binding energy (eV) per atom (E$_b$), the highest (f$_{max}$) and lowest (f$_{min}$) vibrational frequency (cm$^{-1}$) and Hirshfeld Charges ($|$e$|$) of Au$^1$ and Au$^2$. 
     \end{tablenotes}
\end{threeparttable}
\end{table*}

We have calculated some of its isomers in different forms to compare properties of this structure and the comparison results are shown in Table 2. The core-shell structure with Ih symmetry has lower binding energy. That is similar to the predictions and speculations of Juarez L. F. Da Silva et al. in 2010, the energy dropped by the Au diffusion from the core region to the surface, which is driven by surface compression (only 12 atoms) on the core region\citep{34}. The above simulation of 500K also bears it out. As shown in Table 2, the core -shell structure with Ih symmetry has a higher HOMO-LUMO gap, that is, it has less chemical activity than some of its isomers. In other words, this structure has better chemical stability relative to its isomers and this nature also gives this structure a wider application space.

  \begin{table}[!ht]
  \caption{The comparison of some isomers of Au$_{42}$.\tnote{1}}
   \centering
   \renewcommand\tabcolsep{2.5pt}
   \begin{threeparttable}
     \begin{tabular}{*5{c}}\toprule
        & core-shell & core-shell(1) & cage-like & close-packed\\ \midrule
       Sym & Ih	 & Cs	& Ih & C2  \\
       Gap & 0.624 eV & 	0.213 eV	 & 0.425 eV	& 0.063 eV  \\
       E$_{b}$ & 2.23 eV & 	2.46 eV  & 	2.28 eV	& 2.29 eV \\
 \bottomrule
     \end{tabular}
     \begin{tablenotes}
       \footnotesize
       \item[1] Four structures were calculated here, include the core-shell structure with Ih symmetry(shown in Figure 1), the cage-like cluster with icosahedron structure, the core-shell cluster from the Cambridge Cluster Database, and a close-packed structure of Au$_{42}$ with hcp. The data in this table contain their symmetry(Sym), the HOMO-LUMO gap(Gap), and the binding energy(E$_{b}$), the functional used is PBE.
     \end{tablenotes}
   \end{threeparttable}
  \end{table}

The spin magnetic properties and the atomic charge have also been investigated via Hirshfeld analysis (the detailed results are in Table S2). Each atom on the inner core has 0.0174$|$e$|$ positive charges on average and each atom on the shell has 0.0353$|$e$|$ negative charges on average. It turns out that there is a charge transfer between the core and shell. As shown in the electron density and the differential electron density map (Figure 4), there is an almost uniform electron shell (in Figure 4a) and the electrons locate mainly the centers of 12 pentagons on the shell (in Figure 4b), the electron on the inner core flow to the outer shell. Besides, it may be clearly explained by the multi-center bonds between the core and shell, which will be analyzed in detail in the subsequent part.

\begin{figure}[h!]
\centering
\includegraphics[scale = 0.25]{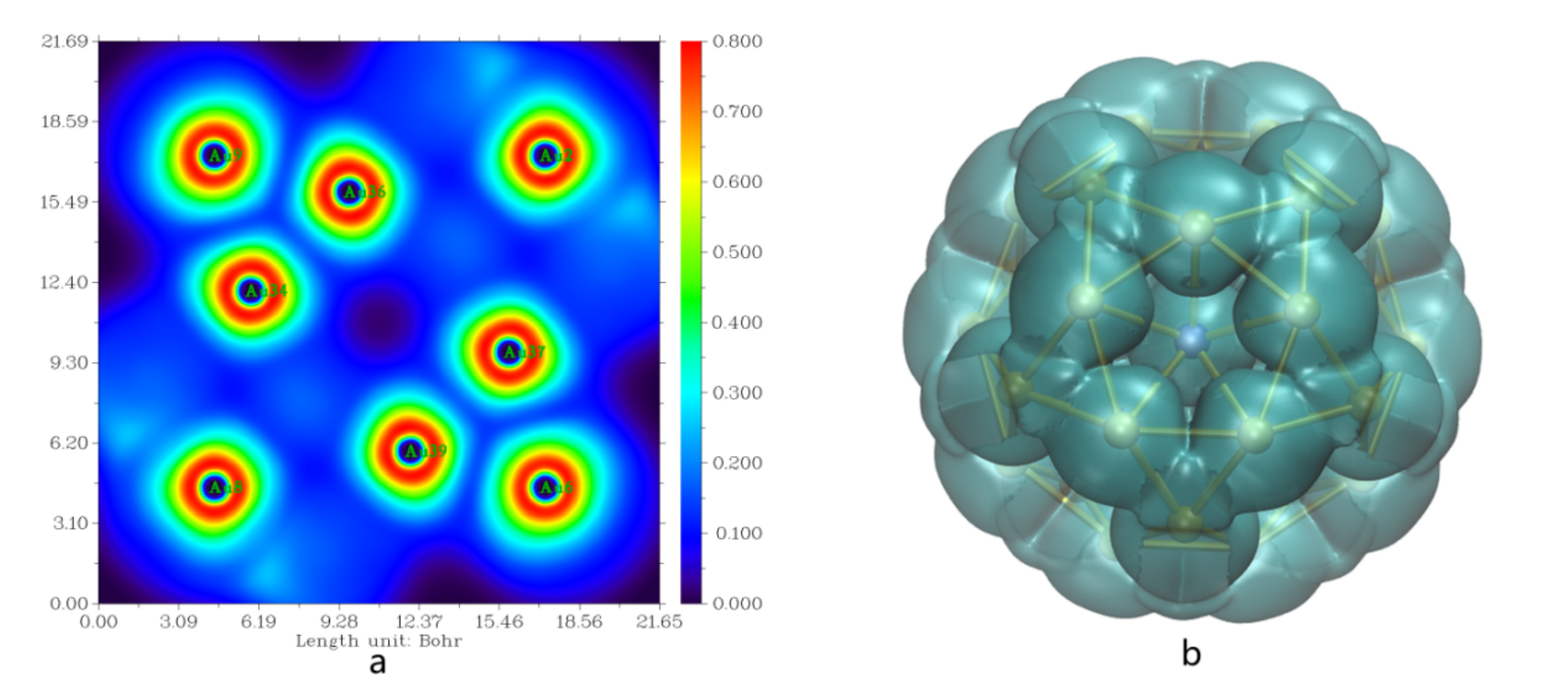}
\caption{The two-dimensional (a) and three-dimensional (b) electronic local functions, the isosurface value is 0.017 e/Å$^3$ in (b).}
\label{03}
\end{figure}

\begin{figure}[h!]
\centering
\includegraphics[scale = 0.2]{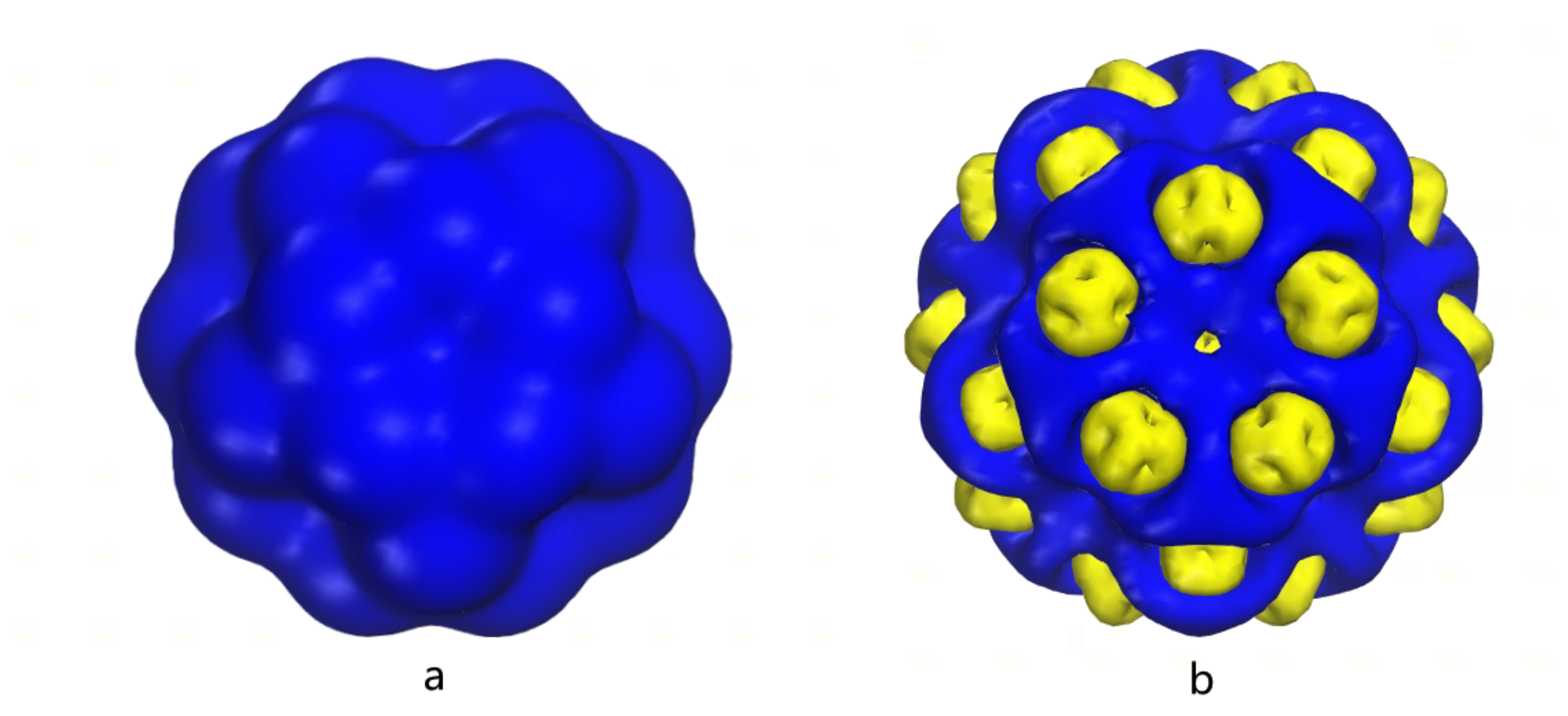}
\caption{(a) Electron density (isosurface value is 0.017 e/Å$^3$) and (b) electron deformation density (isosurface value is 0.017 e/Å$^3$) for the core-shell structure. The blue part shows the charge accumulation and the yellow part shows the charge dissipation.}
\label{04}
\end{figure}

The analyses of the AdNDP reveal the bonding characteristics. For the Au$_{42}$ neutral molecule, there are 462 valence electrons in all. Of the 462 valence electrons in this core-shell structure, there are 210 onecenter lone electron pairs, 20 delocalized six-center two-electron (6c-2e) bonds between the core and shell (shown in Figure 5(a,b)), one delocalized 12-center two-electron (12c-2e) bonds in the core (shown in Figure 5(c,d)). Two kinds of multi-center bonds are both $\sigma$ bonds, and their occupation numbers are 1.86 and 1.93 respectively.

\begin{figure}[h!]
\centering
\includegraphics[scale = 0.4]{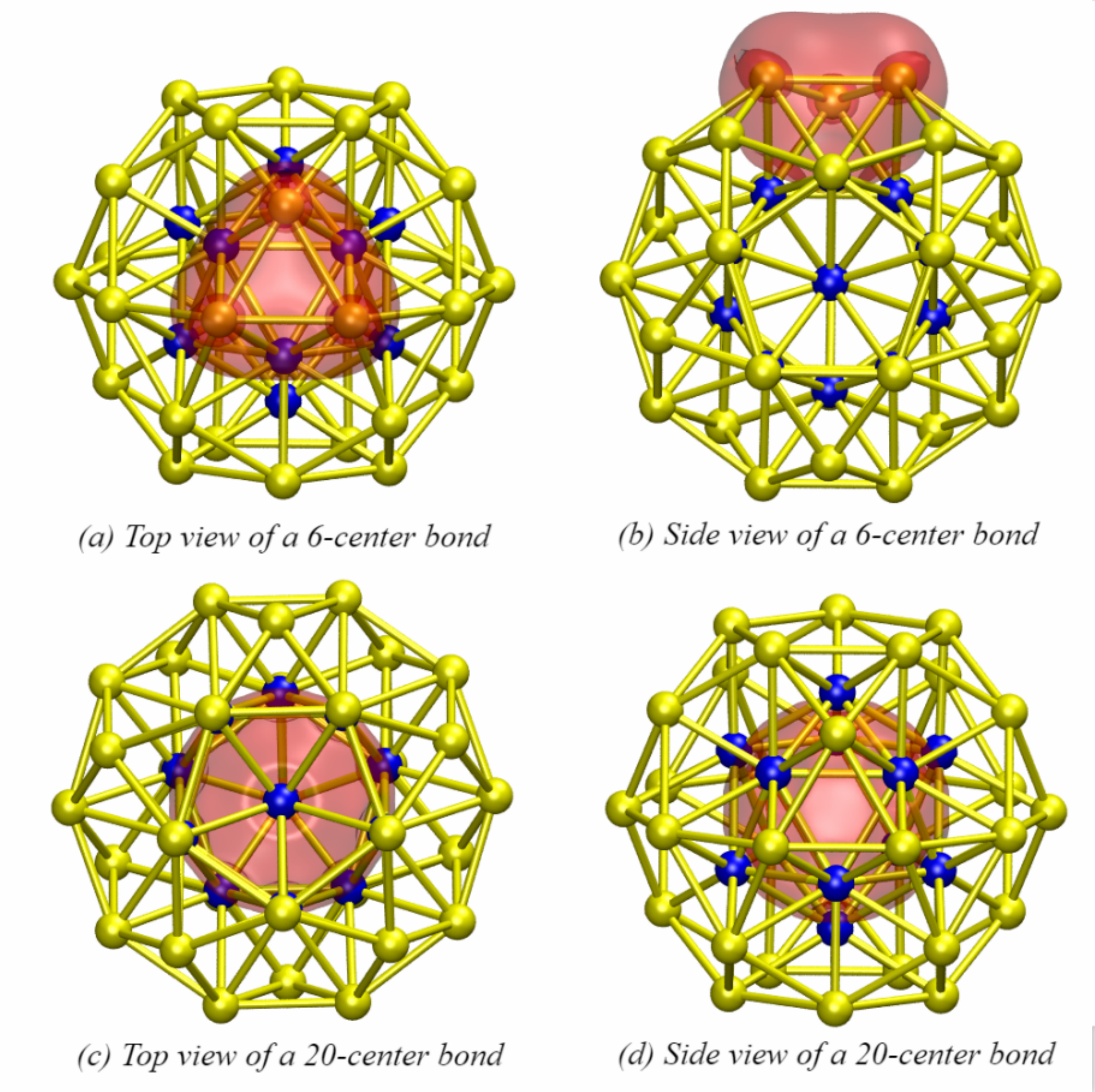}
\caption{Top view and side view of the 6c-2e and 12c-2e  bonds with the isosurface value of 0.02 e/Å$^3$. The 6c-2e bond is composed of three Au$^1$ and three Au$^2$, and the 12c-2e bond is composed of 12 (that is, all) Au$^2$ atoms.}
\label{05}
\end{figure}

For the 6c-2e delocalized bond, each shell atom Au$^1$ provides 18.53$\%$ of the total, while the contribution of each Au$^2$ is about 14.81$\%$. The detailed analysis shows that the constituent from Au$^1$ to the 6c-2e bond is mainly from 6s and 6p electrons and the proportion is about 4.3:1. The contribution of Au$^2$ to it is also mainly 6s and 6p electrons, but the ratio is about 1:3.4. For the 12c-2e bond, the ingredients from each Au$^2$ atom to it are mainly from 6s and 6p but the ratio is about 1:2.4 (detailed analysis data of orbital components are shown in Table S2 to Table S5). The above Hirshfeld charge population analysis shows that there is a charge transfer between the core and shell and this transfer is formally realized through the 6c-2e bonds. As shown in Figure 5, the 6c-2e bond in this structure is composed of three shell Au$^1$ and three core Au$^2$. And the spatial distribution is not uniform (as shown in Figure 5(b)), that is, the distribution of the electron cloud is biased towards the shell. This uneven spatial distribution of 6c-2e delocalized bonds leads to the above-mentioned molecular dynamics simulation results (That is, Au$^2$ atoms move toward the shell at a higher temperature). Figure 3 shows the two-dimensional (a) and three-dimensional (b) ELF maps. It may be seen from the sliced plane map Figure 3(a) that most electrons withdraw to the shell atoms (Au$_{30}$), thus yielding the largest value near shell atoms (bottom left and top right). What’s more, the maximum value is only 0.319, so the electron on this cross-section is delocalized, which is exactly in line with the nature of the metal bond. Figure 3(b) can also better exhibit relatively high charge density on the shell. So, the probability of bonding electrons appearing near the shell is relatively bigger, the outer atoms are slightly negatively charged and the inner atoms are slightly positively charged.

From Table 1, shell atoms have a longer bond length than this between core atoms and this phenomenon also can be explicated by the AdNDP analysis. There are two kinds of multi-center bonds in this structure, 6c-2e and 12c-2e. The existence of 6c-2e implicates dshell $>$ dcore(6c-2e is more biased towards the shell) and it’s precisely that the presence of the 12c-2e ties 12 core atoms further together. The presence of two multi-center bonds makes the bond length relationship mentioned above.

\begin{figure}[h!]
\centering
\includegraphics[scale = 0.25]{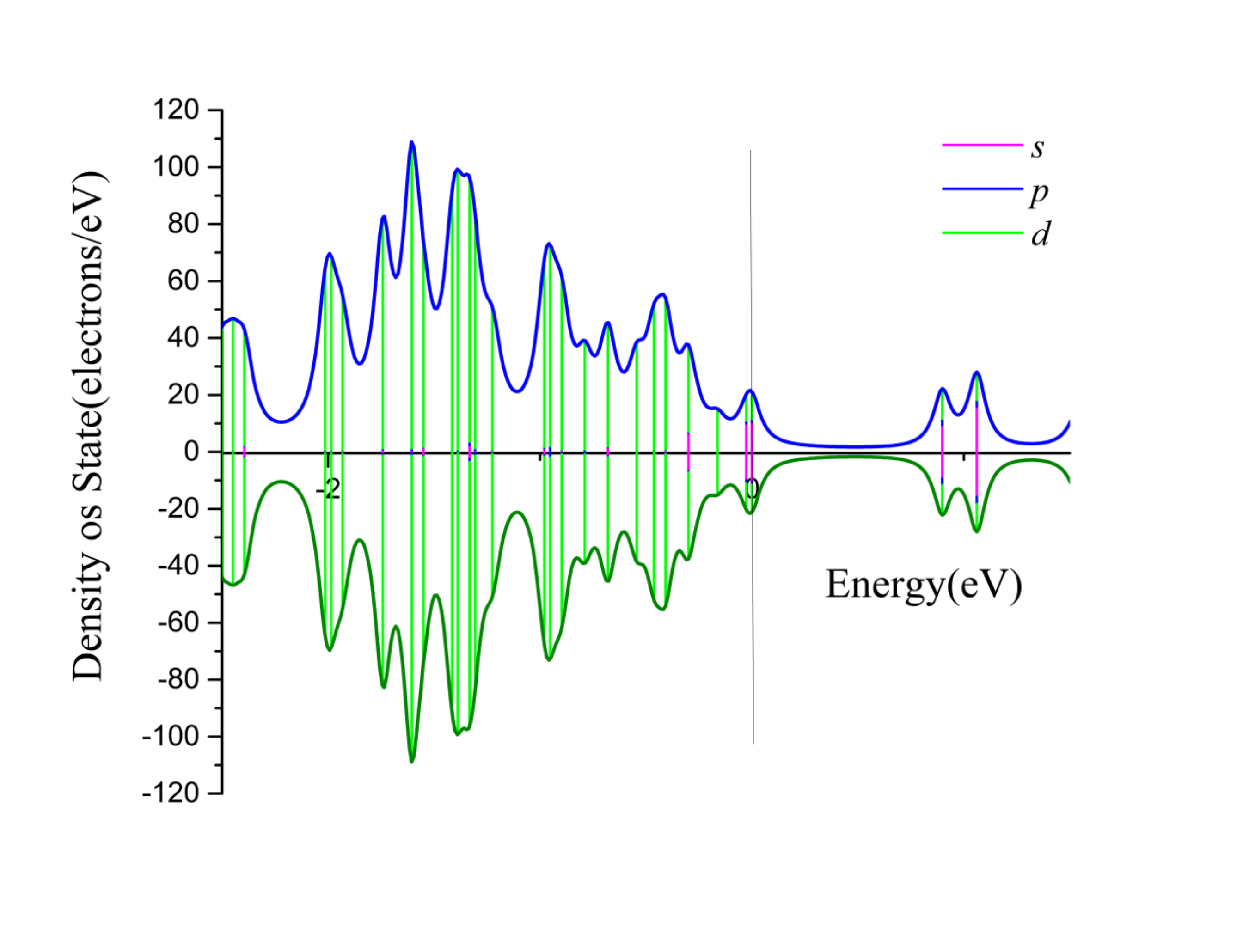}
\caption{PDOS of the core-shell structure of Au$_{42}$ at the GGA/PBE level, the Fermi level is set as 0eV and represented by a line perpendicular to the horizontal axis, the colored vertical lines indicate the relative magnitudes of the different electron states of some molecular orbitals. }
\label{06}
\end{figure}

The electronic distribution is obtained by the partial density of state (PDOS) for this core-shell structure (shown in Figure 6). There is s-d hybridization between the Au atoms at the Fermi level. According to the Hirshfeld population analysis in Table S6, this system is non-magnetic, this point can be verified by the PDOS diagram symmetric about the horizontal axis. About the HOMO of the molecule, some atomic orbitals similar to 5d$_{xy}$, and 5 appear, as shown in Figure 7 (the detailed analysis data on the composition of HOMO is in Table S7). The analysis of AdNDP shows that these electron pairs with shapes similar to d orbitals are not shared by multiple atoms, and the shape of them is very conducive to coordination with some other electron-deficient systems.

\begin{figure}[h!]
\centering
\includegraphics[scale = 0.27]{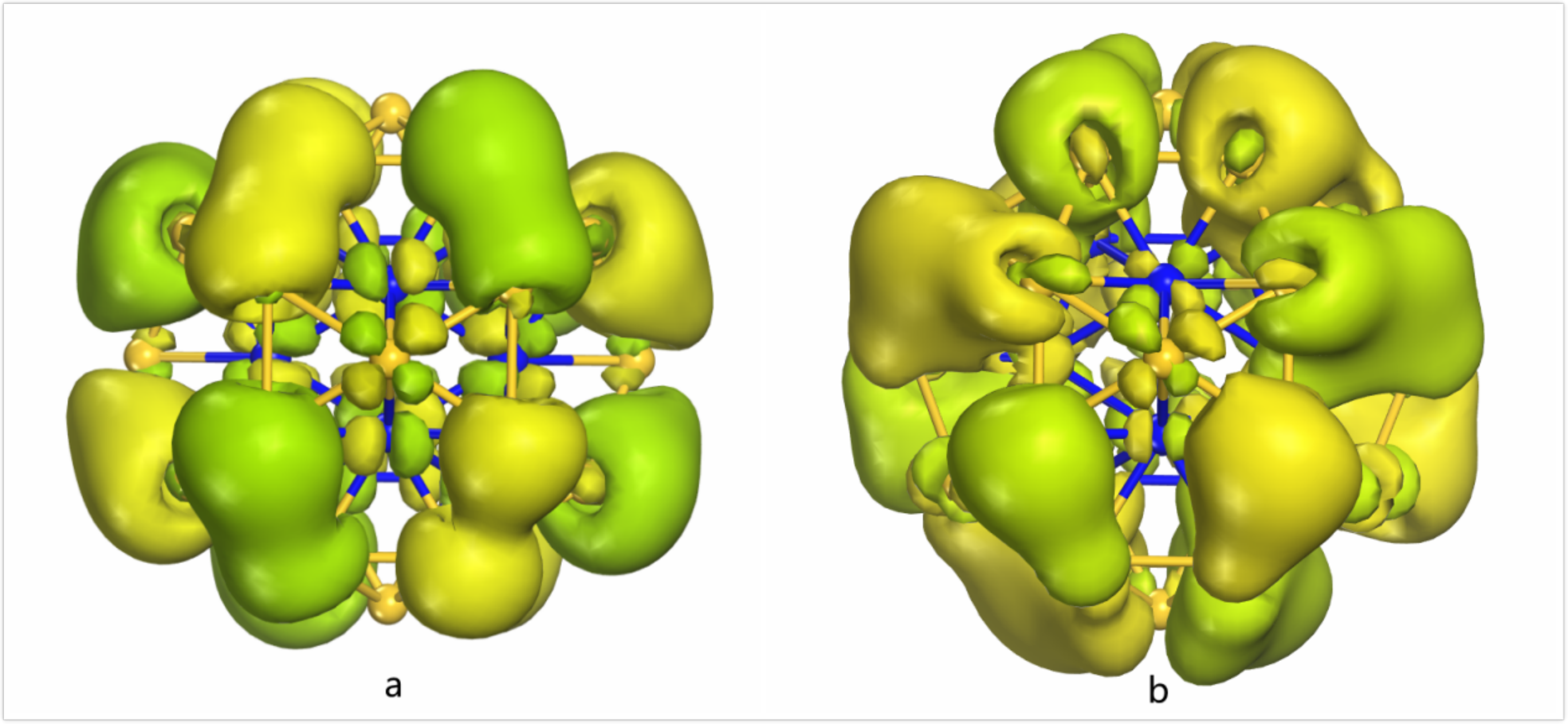}
\caption{The HOMO (a) and LUMO (b) of Au$_{12}$@Au$_{30}$ with the isosurface of 0.017 e/Å$^3$. }
\label{07}
\end{figure}

In summary, we have constructed the Au$_{12}$@Au$_{30}$ from the cage-like icosahedron Au$_{42}$. After energy minimization, it still maintains the core-shell structure with Ih symmetry. By vibrational frequency analysis and molecular dynamics simulations, it is dynamically stable. This structure has a smaller Gap$_{H-L}$ than that of the cage-like Au$_{42}$, indicating its relatively high chemical activity. The AdNDP analyses reveal that there are twenty 6c-2e $\sigma$ bonds and one 12c-2e $\sigma$ bond in this structure and precisely their presence makes the bond length between shell atoms is greater than that of core atoms. The thermodynamic stability and the Hirshfeld charge population can be explained by the 6c-2e delocalized bond joining the core and shell. The PDOS and Hirshfeld charge population analysis showed that the system is closed-shell. The HOMO shows obvious s-d hybridization which may also be verified by the Fermi level of the PDOS. The existence of valence lone electron pairs with obvious d-like orbital characteristics makes the cluster readily coordinate with some electronically deficient systems. This study is of guiding significance to the related experiments for synthesizing pure gold core-shell structures. Because of the large number of lone pairs and higher chemical activity, the highly symmetric core-shell structure may have potential applications in the fields of medicine and catalyst.
\section{Acknowledgement}
This work is supported by the Natural Science Foundation of Hebei Province (E2019105073), the Science Foundation of Hebei Education Department for Young scholar (Grant Nos. QN2019074), and the Ph.D. foundation of Tangshan Normal University (No. 2016A06).
\section{Author Contributions}
Chunyuan Bai: Data curation, Writing-Original draft preparation. Shulan Liu: Resources. Zun Xie: Conceptualization, Supervision, review \& editing. Hongfei Li \&4 Yichen Dong: Validation.
\section{Data Accessability}
The data supporting the results of this study are provided in the supplementary material for this article. If there is still need, the data that support the findings of this study are available from the corresponding author, upon reasonable request.

\bibliography{main}
\bibliographystyle{plainnat}

\end{document}